\documentclass[aps,prb,twocolumn,showpacs,preprintnumbers,amsmath,amssymb,superscriptaddress]{revtex4-1}

\usepackage{graphicx}
\usepackage{dcolumn}
\usepackage{bm}

\begin{document}

\title{On the magnetization textures in NiPd nanostructures}

\author{J.-Y. Chauleau}
\affiliation{Laboratoire de Physique des Solides (LPS), CNRS UMR
8502, Univ. Paris-sud, 91405 Orsay Cedex, France}
\author{B.J. McMorran}
\affiliation{Center for Nanoscale Science and Technology, National
Institute of Standards and Technology (NIST), Gaithersburg, MD
20899-6202, USA}
\author{R. Belkhou}
\affiliation{Synchrotron SOLEIL, l'Orme des merisiers, Saint-Aubin, 91192 Gif-sur-Yvette,
France}
\author{N. Bergeard}
\affiliation{Laboratoire de Physique des Solides (LPS), CNRS UMR
8502, Univ. Paris-sud, 91405 Orsay Cedex, France}
\affiliation{Synchrotron SOLEIL, l'Orme des merisiers, Saint-Aubin,
91192 Gif-sur-Yvette, France}
\author{T.O. Mente{\c{s}}}
\affiliation{ELETTRA, Sincrotrone Trieste S.C.p.A., 34149 Basovizza, Trieste, Italy}
\author{M.{\'{A}} Ni{\~{n}}o}
\altaffiliation[Now at ]{Instituto Madrile{\~{n}}o de Estudios Avanzados en Nanociencia
(IMDEA Nanociencia), Cantoblanco, 28049 Madrid, Spain.}
\affiliation{ELETTRA, Sincrotrone Trieste S.C.p.A., 34149 Basovizza, Trieste, Italy}
\author{A. Locatelli}
\affiliation{ELETTRA, Sincrotrone Trieste S.C.p.A., 34149 Basovizza, Trieste, Italy}
\author{J. Unguris}
\affiliation{Center for Nanoscale Science and Technology, National
Institute of Standards and Technology (NIST), Gaithersburg, MD
20899-6202, USA}
\author{S. Rohart}
\affiliation{Laboratoire de Physique des Solides (LPS), CNRS UMR
8502, Univ. Paris-sud, 91405 Orsay Cedex, France}
\author{J. Miltat}
\affiliation{Laboratoire de Physique des Solides (LPS), CNRS UMR
8502, Univ. Paris-sud, 91405 Orsay Cedex, France}
\affiliation{Center for Nanoscale Science and Technology, National
Institute of Standards and Technology (NIST), Gaithersburg, MD
20899-6202, USA} \affiliation{Maryland NanoCenter, University of
Maryland, College Park, MD 20742, USA}
\author{A. Thiaville}
\affiliation{Laboratoire de Physique des Solides (LPS), CNRS UMR
8502, Univ. Paris-sud, 91405 Orsay Cedex, France}

\date{\today}

\begin{abstract}
We have observed peculiar magnetization textures in
Ni$_{80}$Pd$_{20}$ nanostructures using three different imaging
techniques: magnetic force microscopy, photoemission electron
microscopy under polarized X-ray absorption, and scanning electron
microscopy with polarization analysis. The appearances of
diamond-like domains with strong lateral charges and of weak stripe
structures bring into evidence the presence of both a transverse
and a perpendicular anisotropy in these nanostrips. This anisotropy
is seen to reinforce as temperature decreases, as testified by a
simplified domain structure at 150~K. A thermal stress relaxation
model is proposed to account for these observations. Elastic
calculations coupled to micromagnetic simulations support
qualitatively this model.
\end{abstract}


\maketitle
\section{Introduction}

The nickel-palladium alloys (denoted here as NiPd), which form a
solid solution over the whole concentration range, have been the
subject of many studies for their magnetic
properties.~\cite{Sadron32,Neel32,Crangle65,Ferrando72} Indeed,
palladium is the 4d parent of nickel. It is close to being
ferromagnetic according to the Stoner criterium, and forms a
ferromagnetic alloy with nickel down to Ni atomic concentrations as
small as $\approx 2$~\%,~\cite{Murani74,Kontos04} with a smoothly
varying Curie temperature,~\cite{Hansen58} due to its large magnetic
polarizability.~\cite{Vogel97} Thus, in recent years, NiPd alloys
have been used as ferromagnets of tunable strength for studying the
ferromagnet-superconductor proximity effect.~\cite{Kontos01}
Furthermore, Pd as a noble metal has been shown to provide good
electrical contacts with carbon nanotubes,~\cite{Javey03} a property
retained by the Ni rich phase,~\cite{Sahoo05a} so that
Pd$_{1-x}$Ni$_{x}$ ($x\approx0.7$) has been demonstrated to perform
as a good spin injector and analyzer in the study of gated
spin-transport in carbon nanotubes.~\cite{Sahoo05b,Man06} For such
studies, NiPd occurs in the form of nanostructures, in which the
magnetization orientation is expected to be controlled by the
nanostructure shape and the applied field.

With soft magnetic materials like NiFe, the magnetostatic energy
(the so-called shape anisotropy) gives rise to a preferred
orientation in the direction of the long edge of the nanostructure,
with a coercive field that decreases as the nanostructure width
increases, providing good control of the magnetization in the
magnetic electrodes. For NiPd electrodes, however, this appears not
to be the case. Sahoo et al.~\cite{Sahoo05a} indeed observed, when
applying field along the length of the electrodes, a progressive
magnetization reversal. The switching characteristics changed
completely when applying field in the direction transverse to the
electrodes,~\cite{Feuillet10} as explained by the first magnetic
force microscopy images obtained at LPS that constitute the
starting point of this work. In these two cases, the palladium
atomic concentration was 25~\% to 30~\%. Additionally, anisotropic
magnetoresistance (AMR) measurements performed on electrodes (with
an estimated 40~\% palladium content) showed that the magnetization
was very far from the longitudinal orientation.~\cite{Gonzalez08}
From a comparison of AMR signals measured for fields oriented along
several directions, that study concluded moreover that the
magnetization was, on the average, tilted out of the plane. The
existence of a strong perpendicular anisotropy in infinite films was
also directly confirmed by ferromagnetic resonance
measurements,~\cite{Gonzalez08} and could also be guessed from the
extraordinary Hall effect measurements on 90~\% Pd rich
samples.~\cite{Kontos01} In view of this complexity, we push here
the study one step further by imaging the complex magnetization
textures in the NiPd electrodes, using different magnetic imaging
techniques with a high spatial resolution, namely magnetic force
microscopy (MFM), photoemission electron microscopy combined with
X-ray magnetic circular dichroism (XMCD-PEEM) and scanning electron
microscopy with polarization analysis (SEMPA). The possibilities and
characteristics of these techniques are indeed
complementary:~\cite{Hopster05} we used MFM to get a global image of
the structure with no depth or component resolution, XMCD-PEEM to
probe the surface magnetization componentwise and also at low
temperature, and SEMPA for vectorial maps of the surface
magnetization. In order to interpret quantitatively the results
obtained, an analysis based on the differential thermal expansion of
film and substrate, including elastic and micromagnetic simulations,
is proposed and discussed as a cause for the observed anisotropies.


\section{Imaging and qualitative analysis}
The structures under study are Ni-rich NiPd nanostrips with varying
widths $w$ from 100~nm to 1000~nm and thicknesses $t$ between 10~nm
and 50~nm, the length being 5~$\mu$m unless otherwise specified,
with a 3~nm Pd or Al cap to protect against oxidation. They have
been patterned using a lift-off technique, by e-beam lithography and
e-gun UHV evaporation with deposition rate around $0.13$~nm/s, onto
Si substrates with native oxide. The saturation magnetization
($M_S\approx 3.2\times10^{5}$~A/m) and the typical composition have
been measured by, respectively, alternating gradient force
magnetometry (AGFM) and Rutherford back-scattering spectroscopy
(RBS), giving an atomic composition of Pd of $\approx 20~\%$
(slightly drifting with source usage). Note that the following
results involve only the virgin magnetization states, but that a
magnetic field has also been applied, showing that the magnetization
textures under study are more robust than simple metastable states
generated during growth.

\begin{figure}[!h]
\includegraphics[width=0.7\columnwidth]{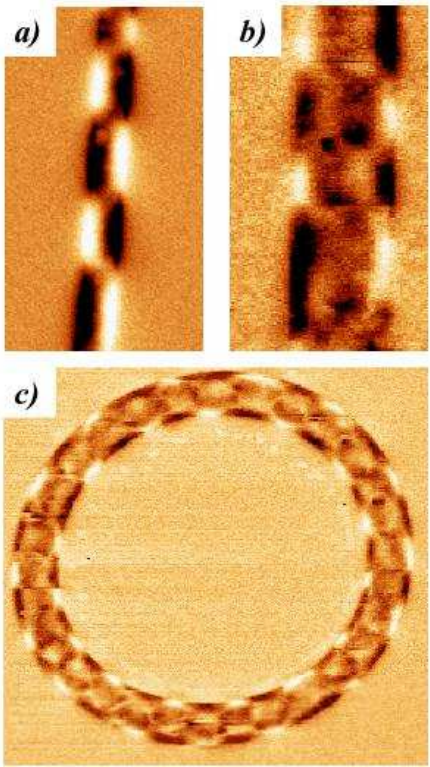}
\caption{(color online) Selection of MFM images on different NiPd nanostructures:
(a) $150$~nm wide and $30$~nm thick nanostrip;
(b) $450$~nm wide and $30$~nm thick nanostrip;
(c) NiPd nanorings with a diameter of $5~\mu$m, a width of $500$~nm and a thickness of $30$~nm.}
\label{fig:MFM}
\end{figure}%

The MFM contrast of $30$~nm thick narrow strips (Fig.~\ref{fig:MFM}a) reveals
alternate edge magnetic charges.
A clear correlation between the two sides is also observed,
with magnetic charges on one side facing opposite charges on the other.
This rather uniform pattern corresponds to magnetic domains with a transverse magnetization,
an orientation orthogonal to the longitudinal direction that minimizes the
magnetostatic energy (shape anisotropy).
For slightly wider strips (Fig.~\ref{fig:MFM}b), in addition to the edge magnetic
charges, an inner contrast appears.
This means that the magnetization does not fully lie along the transverse axis but potentially
curls inside, revealing a more complex texture (also potentially perturbed by the stray
field of the MFM tip).
The generality of the transverse orientation is directly attested by the image of a
ring-shape sample (Fig.~\ref{fig:MFM}c).
Note also that no such magnetic structures were observed by MFM on the control unpatterned films.
As MFM probes only the sample magnetic stray field (and on one side of the sample),
magnetization distribution reconstruction from a MFM image is not unique.
Therefore we used two other direct imaging techniques, namely XMCD-PEEM and SEMPA.
Both techniques yield images of magnetization within a few nanometers
from the surface.

\begin{figure}[!h]
\includegraphics[width=\columnwidth]{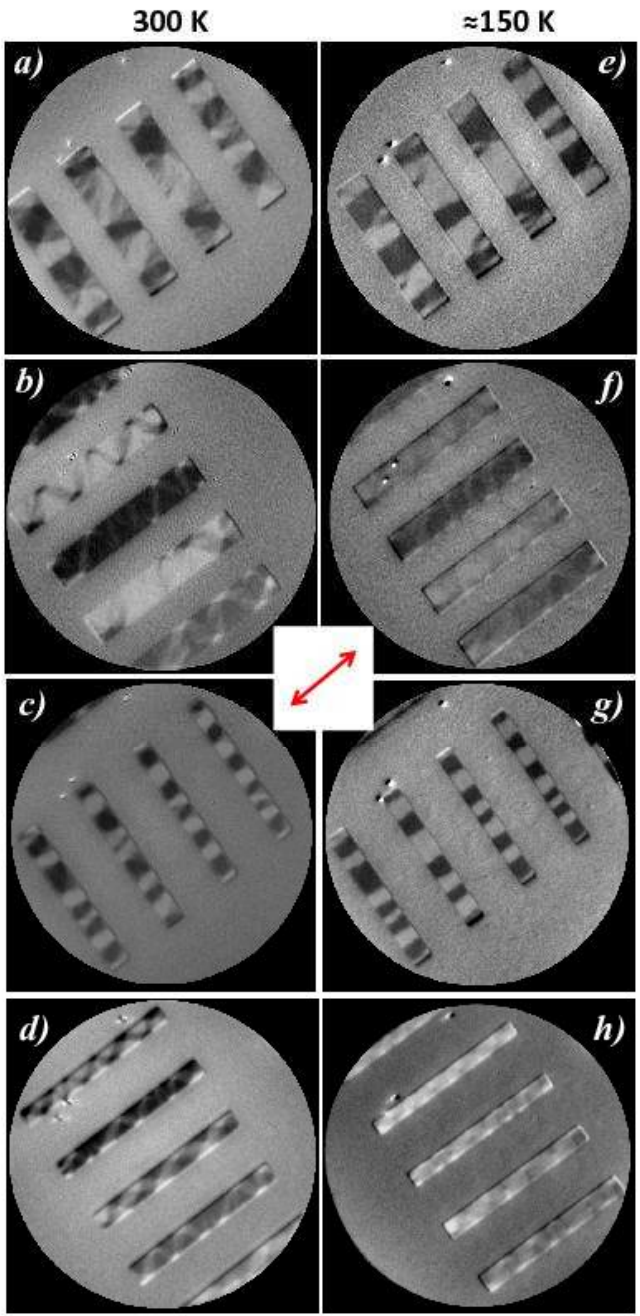}
\caption{XMCD-PEEM images of 4 sets of 5~$\mu$m long and 30~nm thick structures,
taken at room temperature (left column) and at low temperature ($\approx150$~K, right column).
Two series of widths are shown, namely $1~\mu$m to $0.85~\mu$m (a,b,e,f) and
$0.6~\mu$m to $0.45~\mu$m (c,d,g,h), both with a 50~nm step.
Depending on the orientation of the nanostrips with respect to the trace of the X-ray incidence
plane (shown by the double arrow), the transverse (a,c,e,g) or longitudinal (b,d,f,h)
magnetization components are probed (as these two orientations correspond to different
structures, the transverse and longitudinal images for the same nanostructure width can
however not be combined).
}
\label{fig:XPEEM}
\end{figure}%

The XMCD-PEEM experiments were carried out with the combined
PEEM-LEEM (low energy electron microscope) apparatus
\cite{Locatelli06} operating at the Nanospectroscopy beamline of the
Elettra synchrotron, the X-rays being tuned to the $L_3$ edge of Ni.
In the setup used, the circularly polarized X-rays impinge on the
sample at a $16^\circ$ angle from the surface. The differential
absorption of the X-rays (circular dichroism), proportional to the
dot product of magnetization and photon wavevector, therefore
predominantly originates from the in-plane magnetization components,
with a small contribution from the out-of-plane component. The
magnetization images are obtained by forming the difference of PEEM
images acquired with opposite helicity of the X-rays. This method is
inherently surface-sensitive due to the limited electron escape
depth, which is a few nanometers for the typical 2~eV energy of the
collected electrons. We show in Fig.~\ref{fig:XPEEM} only the large
(from $1~\mu$m to $0.85~\mu$m) and intermediate (from $0.6~\mu$m to
$0.45~\mu$m) width nanostructures, for the medium thickness
($30$~nm). Orthogonal sets of strips have been patterned, allowing
us to probe on the same sample the transverse
(Figs.~\ref{fig:XPEEM}a and \ref{fig:XPEEM}c) and the longitudinal
(Figs.~\ref{fig:XPEEM}b and \ref{fig:XPEEM}d) components of
magnetization, albeit on different structures. These magnetic images
corroborate the conclusions drawn from the MFM images, as
Figs.~\ref{fig:XPEEM}a and \ref{fig:XPEEM}c show strong transverse
components, with a higher complexity for the wider structures. On
the other hand, a magnetic contrast is also present for images in
the longitudinal configuration
(Fig.~\ref{fig:XPEEM}b,~\ref{fig:XPEEM}d), proving that the
magnetization is not fully transverse. For intermediate width
(Fig.~\ref{fig:XPEEM}d) as well as narrower structures, symmetric
diamond patterns are observed with no global longitudinal moment.
However, deformed diamond patterns also appear (e.g. for 500~nm and
600~nm width in Fig.~\ref{fig:XPEEM}d), where closure domains on the
long edges with one (longitudinal) magnetization are bigger than for
the opposite magnetization, meaning that such structures have a
non-zero longitudinal moment. For the large widths
(Fig.~\ref{fig:XPEEM}b), this deformation is general, and very
pronounced. Note however that the close proximity of the structures
introduces a dipolar coupling, stabilizing a staggered (between
successive nanostructures) longitudinal magnetization structure,
quite apparent on Fig.~\ref{fig:XPEEM}b by the alternation of bright
and dark overall contrasts. This dipolar coupling gives rise to a
(staggered) applied field along the longitudinal direction.

At this point, magnetization vector maps of the structures are needed.
With XMCD-PEEM, this requires an azimuthal sample rotation and accurate image
matching.
Instead, this is achievable using SEMPA, whereby images of the magnetization direction 
are obtained by measuring the spin polarization of secondary
electrons emitted in the SEM.
Two vector components (either the two in-plane, or one in-plane and the out of plane component) of
the surface magnetization along with the conventional SEM image are acquired simultaneously using
a quadrant spin detector.
For these experiments, performed on the SEMPA at NIST, the sample surface was cleaned in situ by
sputtering with 1~keV Ar ions (checked by Auger spectroscopy) and then capped with a few
monolayers Fe for contrast enhancement.
All images discussed in the following are maps of the surface magnetization distribution resolved
along the in-plane $x$ (longitudinal) and $y$ (tranverse) axes.

\begin{figure}[!h]
\includegraphics[width=\columnwidth]{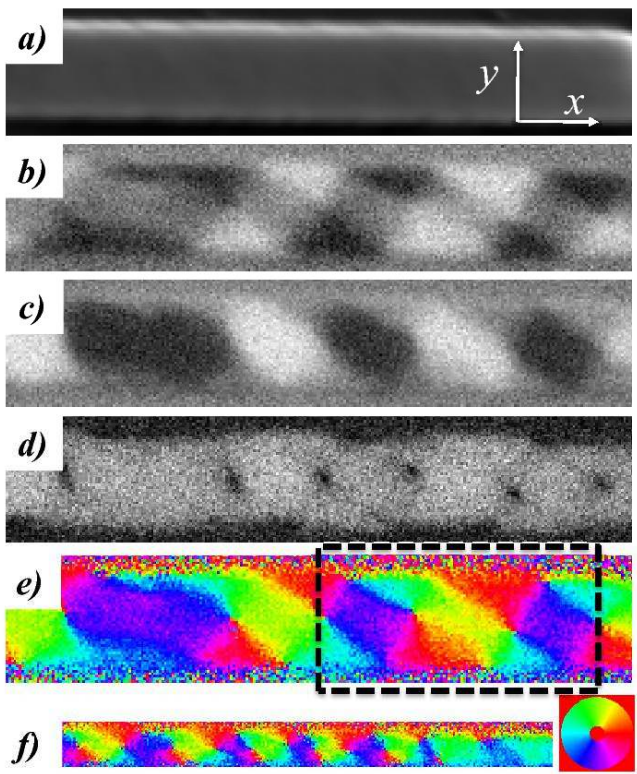}
\caption{(color online) SEMPA imaging of 0.3~${\mu}$m wide and
$50$~nm thick NiPd nanostrip: (a) SEM image; (b) and (c) $x$ and $y$
magnetization component distributions, respectively; (d) in-plane
magnetization magnitude, built from (b) and (c); (e) color-coded
reconstruction of the in-plane magnetization texture (see color
wheel for the direction coding). Note the image distortion due to
drift while scanning. In addition, the in-plane magnetization for a
0.1~${\mu}$m wide nanostrip is shown in (f).} \label{fig:SEMPA}
\end{figure}

\begin{figure}[!h]
\includegraphics[width=\columnwidth]{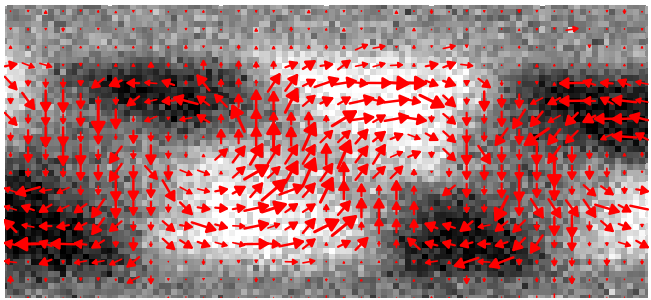}
\caption{(color online) Longitudinal SEMPA image overlaid with the reconstructed magnetization
texture (in-plane components) for the area indicated by the dashed box in Fig~\ref{fig:SEMPA}.
}
\label{fig:SEMPA_reconst}
\end{figure}

Fig.~\ref{fig:SEMPA}a displays a conventional SEM image of a
$0.3~\mu$m wide, $50$~nm thick and 10~$\mu$m long nanostrip,
attesting to the nanopatterning quality. The simultaneously measured
in-plane components of the surface magnetization are shown in
Figs.~\ref{fig:SEMPA}b and \ref{fig:SEMPA}c, featuring the $M_x$ and
$M_y$ distributions, respectively: the longitudinal image
(Fig.~\ref{fig:SEMPA}b) shows the `closure' domains, and the
transverse image (Fig.~\ref{fig:SEMPA}c) shows the `diamond' domains.
Note that there is about 35~nm of longitudinal drift during the
image scan, which adds to the slanted shape of domains.
Fig.~\ref{fig:SEMPA}d is a composite image displaying $M_x^2+M_y^2$,
the magnitude of the in-plane magnetization. The contrast appears
uniform in the image, with a noise level identical to that of the
images of the two components, except at points roughly located along
the strip axis. The latter clearly identify with the vortex cores,
also noticeable in MFM images, yet indirectly. The alternate
off-centered position of the vortices that is very visible in
Fig.~\ref{fig:SEMPA}d is another indication of a non-zero
longitudinal moment of the structure. Combining data in
Figs.~\ref{fig:SEMPA}b and \ref{fig:SEMPA}c allows for a vector
representation of the in-plane magnetization
(Fig.~\ref{fig:SEMPA}e). A magnified vector map superposed onto the
longitudinal contrast is also shown in Fig.~\ref{fig:SEMPA_reconst}.
Similar features are observed on $0.1~{\mu}m$ wide strips
(Fig.~\ref{fig:SEMPA}f).

In addition, we have observed that this transverse behaviour changes
strongly with temperature, by comparing XMCD-PEEM images at room
temperature (Figs.~\ref{fig:XPEEM}a to \ref{fig:XPEEM}d) and at low
temperature ($\approx150~K$, Figs.~\ref{fig:XPEEM}e to
\ref{fig:XPEEM}h) on the same structures. First, comparing the
transverse component (Figs.~\ref{fig:XPEEM}a vs. \ref{fig:XPEEM}e,
and \ref{fig:XPEEM}c vs. \ref{fig:XPEEM}g), it is clear that the
magnetic texture has been simplified when lowering temperature. The
fact that intermediate grey levels have disappeared proves the
reinforcement of the transverse anisotropy. This is corroborated by
the loss of magnetic contrast while probing the longitudinal
component (Figs.~\ref{fig:XPEEM}b vs. \ref{fig:XPEEM}f, and
\ref{fig:XPEEM}d vs. \ref{fig:XPEEM}h).

\begin{figure}[!h]
\includegraphics[width=\columnwidth]{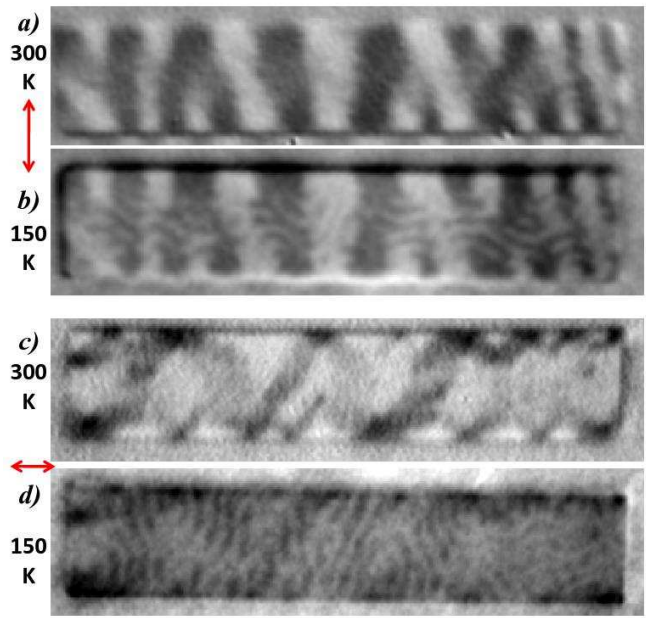}
\caption{XMCD-PEEM images of two 0.9~$\mu$m wide, 50~nm thick NiPd nanostrips,
in the transverse configuration at two temperatures: $300$~K (a) and $150$~K (b) and in the
longitudinal configuration for the same temperatures (c), (d).
Note that the same structure is observed in (a)-(b), and in (c)-(d).
}
\label{fig:XPEEM_900}
\end{figure}

\begin{figure}[!h]
\includegraphics[width=0.8\columnwidth]{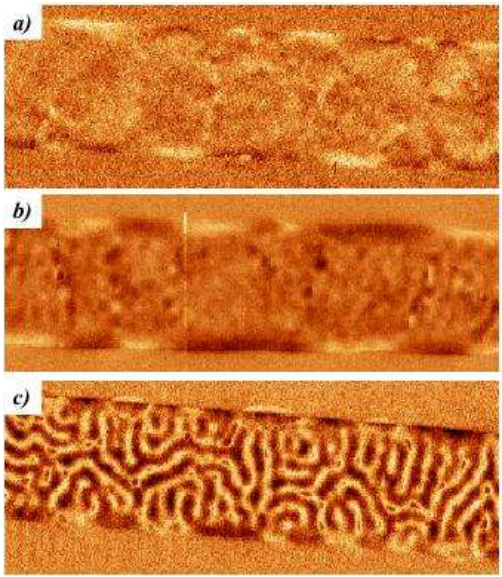}
\caption{(color online) MFM images of a 1~${\mu}$m wide NiPd
nanostrip, with different thicknesses: (a) $30$~nm, (b) $40$~nm and
(c) $50$~nm, showing the appearance of the weak stripe domain
structure. In (b), the largest internal contrast appears along the
domain walls between transverse domains, and is linked to the
wall structure.}
\label{fig:WS_MFM}
\end{figure}

Moreover, at low temperature or for thicker structures ($50$~nm), an
additional contrast with a fine scale appears both in the
longitudinal and transverse configurations (Fig~\ref{fig:XPEEM_900};
a similar contrast was also observed in SEMPA for the 50~nm
structures at room temperature). This is the signature of a
so-called weak stripe domain structure \cite{Hubert98-WS}, a
fingerprint of an additional anisotropy with out of plane easy axis,
with a magnitude that is smaller than the perpendicular
demagnetization energy. In weak stripes, the magnetization at film
center periodically tilts (less than $90^\circ$) out of the plane,
in order to decrease the perpendicular anisotropy energy. The
appearance of surface charges due to this perpendicular component is
avoided by creating periodic `rolls' for the magnetization
components that are transverse to the main (in-plane) magnetization.
Thus, the surface magnetization is mainly in-plane, with a periodic
partial rotation towards the direction transverse to the stripe
elongation. The wavevector of these periodic rotations is orthogonal
to the main magnetization direction in order to avoid magnetic
charges. In top view, weak stripes appear as domains running
parallel to the main magnetization, with an oscillating transverse
magnetization component (as well as with a smaller oscillating
out-of-plane component with 90$^\circ$ dephasing). These weak
stripes provide information about the magnetic structure and
energetics of the sample. (i) As XMCD-PEEM essentially probes the
in-plane magnetization component, the fluctuations due to the weak
stripe structure reveal the mostly longitudinal stripes for
transverse incidence and vice-versa. Thus, the observation of
longitudinal weak stripes in the nanostrip center
(Fig.~\ref{fig:XPEEM_900}b), and of transverse weak stripes over all
the sample width, with a reinforcement of their contrast at the long
edges of the structures (Fig.~\ref{fig:XPEEM_900}d), shows that the
magnetization is globally transverse, with longitudinal components
that are largest around mid-width. (ii) Even if weak stripes are
barely observable at room temperature using XMCD-PEEM or SEMPA (not
shown), MFM proves that they are already present
(Fig.~\ref{fig:WS_MFM}) when the thickness is large enough. This
feature is traced back to the surface sensitivity of XPEEM or SEMPA,
compared to the volume sensitivity of MFM. Thus, the critical
thickness $D_{cr}$ at which weak stripes appear can be estimated
from the MFM images. Fig.~\ref{fig:WS_MFM} indeed shows that,
whereas at 50~nm the stripe pattern is well established, at 30~nm
nothing is observed, and at 40~nm some modulation is visible at some
places, so that $D_{cr} \approx 50$~nm. This value will be later
compared to calculations. (iii) Fig.~\ref{fig:WS_MFM}c indicates
that the transverse anisotropy dominates close to the edges since
weak stripes meet the edges at right angles. Note that at these
nanostrip dimensions, it is still possible to observe the bright and
dark edge contrast, yet with no correlation anymore between the two
sides. (iv) The fact that the surface magnetic contrast of the weak
stripes increases at low temperatures reveals an increase of
perpendicular anisotropy with respect to the demagnetizing energy.

To sum up these observations, a transverse anisotropy, competing with demagnetising energy,
has to be invoked in order to maintain a stable diamond-like magnetic structure.
In addition, we noted the presence of perpendicular anisotropy sufficient to allow for a
weak stripe pattern to appear for $50$~nm thick nanostrips.
Both contributions are enhanced when temperature is decreased.


\section{Quantitative analysis}

We now discuss the origin of both transverse and perpendicular
anisotropies. First, we note that the orientation of the long axis
of the strips, relative either to the substrate or a deposition
angle, has no noticeable effect on the observed magnetization
texture. Considering the out-of-plane and edge-localised transverse
anisotropies, the evolution of the magnetic contrast with
temperature, and the independence of the effect on the orientation
of the nanostructures (Fig.~\ref{fig:MFM}c), thermal stresses are a
likely cause to all these phenomena. Indeed, metals and
semiconductors have very different thermal expansion coefficients
$\alpha$, for example $\alpha_{Ni} = 13.3\times10^{-6}$~K$^{-1}$ and
$\alpha_{Si} = 2.49\times10^{-6}$~K$^{-1}$. Assuming that the
temperature during layer deposition is higher than room temperature
(the sample holder was not cooled), a thermal stress exists in the
NiPd layer at room temperature, that is reinforced at low
temperature ($\Delta T = T_\mathrm{growth} -
T_\mathrm{observation}$).

\begin{figure}[!h]
\includegraphics[width=\columnwidth]{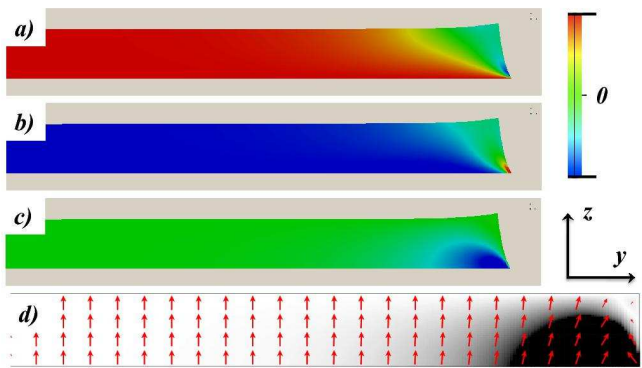}
\caption{(color online) End-view cross-section representation of the thermal strain
components in a  30~nm thick and 500~nm wide nanostrip
calculated for an interfacial strain $\epsilon_0=10^{-3}$ (corresponding to $\Delta T= 92.5$~K
for nickel), for an infinitely rigid substrate (for the meaning of axes, compare
with Fig.~\ref{fig:SEMPA}).
The displacements have been enhanced by a factor of $100$ in order to be observable.
The strain components are:
(a) the transverse strain $\epsilon_{yy}$; (b) the perpendicular strain $\epsilon_{zz}$
and (c) the shear strain $\epsilon_{yz}$.
The color scale extends between $-\epsilon_0$ and $\epsilon_0$ for $\epsilon_{yy}$
and between $-\epsilon_1$ and $\epsilon_1$ for $\epsilon_{zz}$ and $\epsilon_{yz}$.
A half cross-section is represented because of symmetry.
The anisotropy distribution is shown in (d), with arrows giving the easy axis direction and
the grey levels coding the value of the transverse anisotropy, the color code clipping
values (evaluated with the nickel parameters) above 3~kJ/m$^3$.}
\label{fig:elas}
\end{figure}

In order to evaluate quantitatively this effect, the elastic problem
in 2D (infinite nanostrip length) was numerically solved, for
different nanostrip transverse dimensions. 
This was performed with a
homemade finite differences code, in the framework of isotropic
elasticity. Interpolated values for the elastic coefficients of NiPd
were adopted, namely Young's modulus $E= 187.7$~GPa and Poisson's
ratio $\nu= 0.35$ \cite{Yamamoto51,Rayne60}. 
Figs.~\ref{fig:elas}a
to \ref{fig:elas}c show the strain components distribution across
the section of a 30~nm thick and 500~nm wide nanostrip, assuming an
infinitely rigid substrate. 
At the top edges, the NiPd is fully
relaxed, noticeable on the $\epsilon_{yy}$ and $\epsilon_{zz}$ maps
by the zero strain regions (see color code). 
Along the cross-section
symmetry plane, the behaviour is the same as that expected for an
infinite film, i.e. $\epsilon_{yy}=\epsilon_0 \equiv \Delta\alpha
\Delta T$ and $\epsilon_{zz}=-2\nu/(1-\nu)\epsilon_0 \equiv
-\epsilon_1$.
Here, $\epsilon_0$ is the initial interfacial strain,
i.e. the product of the thermal expansion coefficient difference
$\Delta \alpha$ and the temperature difference $\Delta T$, 
and $\epsilon_1$ is the perpendicular strain in
an infinite film (both numbers are defined positive, but as
temperature is lower than that during growth the film is in
compression along the out-of-plane direction and in tension
in-plane). 
Besides, one can note a very localised non-zero shear
strain at the bottom edges of the nanostrip (Fig.~\ref{fig:elas}c),
that corresponds to the inclination of the lateral edge of the
nanostrip.

Once the strains are known, considering the isotropic magnetoelastic coupling with a
negative magnetostriction
coefficient ($\lambda_S=-36\times10^{-6}$ for Ni), it is possible to compute an
anisotropy distribution map (Fig.~\ref{fig:elas}d).
Note that, because of the existence of shear, the magnetoelastic energy must be diagonalised
at every location in order to get the local easy axis direction and anisotropy constants.
Since $\lambda_{S}$ is negative, the easy axis lies along the compression axis, i.e.
perpendicular to the sample plane in the middle (Fig.~\ref{fig:elas}d) as already
measured for Ni rich films.~\cite{BenYoussef04}
The finite size of the strips allows strain relaxation at the edges with the
appearance of shearing, driving the easy axis to locally rotate towards the
transverse direction.
Therefore, thermal strains do lead to out-of-plane and transverse anisotropies, that
vary with position, the transverse anisotropy being a `side-effect' of the
perpendicular anisotropy.
In addition, the average value of the latter will depend on the aspect ratio of the
strip cross-section since it rests on edge contributions.
This dependence is supported by indirect magnetization measurements of some nanostrips
by anisotropic magnetoresistance measurements (not shown).

\begin{figure}
\includegraphics[width=\columnwidth]{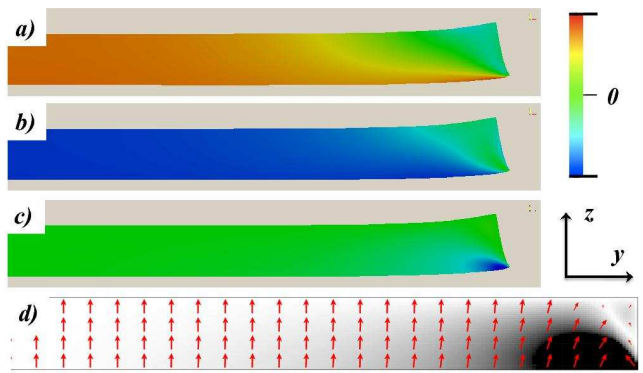}
\caption{(color online) Same data as in Fig.~\ref{fig:elas}, but computed for a deformable
Si substrate (treated as an isotropic medium with $E=185$~GPa and $\nu= 0.26$).~\cite{Wortman65}
The substrate was an infinite parallelepiped with 1.5~$\mu$m edge size, with zero displacement
boundary conditions applied at the three sides without structure.}
\label{fig:elas2}
\end{figure}

The elastic calculations were repeated for a deformable substrate, and the results are
shown in Fig.~\ref{fig:elas2}.
The deformations in the NiPd layer are reduced, as one part of the thermal stress is
relaxed in the substrate.
As a result, the induced anisotropy is also reduced, keeping roughly the same distribution.

The next step is to quantify the perpendicular anisotropy constant
$K_{perp}$. The appearance of the weak stripe domains provides
information about this constant since the critical thickness
($D_{cr}$) for this depends on the quality factor
$\mathcal{Q}=2K_{perp}/\mu_0M_S^2$.~\cite{Hubert98-WS,Vukadinovic01}
In the present case, $D_{cr}$ is about $50$~nm. Considering an
exchange constant of $0.6\times10^{-11}$~J/m, we obtain an exchange
length of $9.6$~nm. Then, from results due to Vukadinovic et
al.,~\cite{Vukadinovic01} the quality factor's critical value is
$\mathcal{Q}=0.48$, corresponding to a perpendicular anisotropy
constant $K_{perp}\approx3\times10^4$~J/m$^3$. This value is of the
same order of magnitude as previously obtained by FMR for infinite
films.~\cite{Gonzalez08} The anisotropy can be linked to a
difference of temperature via the following relation (valid for an
infinite film):

\begin{equation}
K_{perp}=\frac{3}{2}.\frac{E\lambda_S\Delta\alpha}{1-\nu}.\Delta T.
\end{equation}
Therefore, $\Delta T\approx 164$~K is needed to reach the required value, when
considering an isotropic elastic constant $E=193.5$~GPa and
$\nu=0.382$ for nickel.~\cite{Yamamoto51} 
However, if we consider
reported values for bulk NiPd with 20~\% Pd, namely $\alpha = 16
\times 10^{-6}$~K$^{-1}, $~\cite{Masumoto70} and $\lambda_S \approx
-44 \times 10^{-6}$,~\cite{Tokunaga78} and use the interpolated
values for the elastic coefficients used in the calculations, we
obtain a lower value $\Delta T \approx 116$~K.

The calculated 2D anisotropy distributions were used as inputs into
the OOMMF software~\cite{OOMMF} to simulate the equilibrium magnetic
structures as a function of nanostrip dimensions. Keeping a 5~$\mu$m
long strip as in the experiments, the simulation started with 11
transverse magnetic domains (as observed), and with a noise of
$10\%$ on the magnetization direction in order to avoid metastable
states. Note that the anisotropy distribution, obtained for an
infinitely long strip by a 2D calculation, is not correct at the $x$
ends of the structures, so that only the internal structures should
be considered. For a $50$~nm thick nanostrip, weak stripe domains
appeared (not shown), as expected.

\begin{figure}[!h]
\includegraphics[width=\columnwidth]{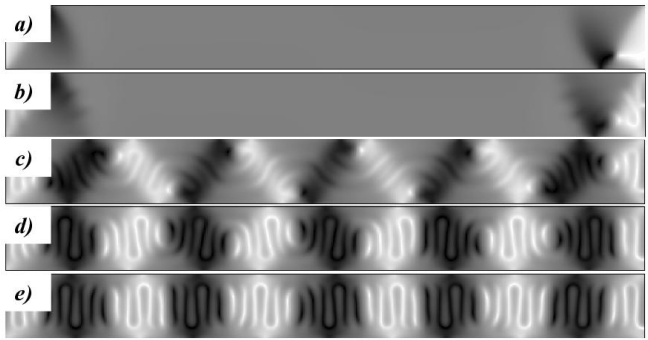}
\caption{Micromagnetic simulations of the equilibrium magnetization
distribution (top view) in a $0.5~\mu$m wide and $30$~nm thick
nanostrip for different values of $\Delta T$ (computed with the parameters of nickel),
using the anisotropy distribution shown in Fig.~\ref{fig:elas}~d: (a) $\Delta T =160$~K,
(b) 250~K, (c) 260~K, (d) 270~K and (e) 280~K.
The mesh size is $5 \times 5 \times 5$~nm$^3$, much below the exchange length
($\Lambda\approx9.6$~nm).
The grey scale represents the magnetization transverse component, at the sample surface.
As the full equilibration of the weak stripe domains (and of the number of transverse domains)
requires an infinite number of iterations, the structures shown are meaningful locally, but
maybe not globally.
}
\label{fig:micromag}
\end{figure}%

Fig.~\ref{fig:micromag} presents the converged
magnetic configurations in a $0.5~\mu$m
wide and $30$~nm thick nanostrip for different values of $\Delta T$ (calculated
using the parameters of nickel; they would be 1.4 times smaller with the
parameters of NiPd considered here).
The calculations assumed the room temperature NiPd parameters; only the magnitude
of the thermal stress induced anisotropy was scaled when changing the temperature
difference $\Delta T$.
Whereas for $\Delta T= 160$~K (a) and up to
$\Delta T=250$~K (b) the $11$ domain initial state eventually becomes fully
longitudinal, from $\Delta T=260$~K (c, d) a domain structure very similar to that
experimentally observed is obtained, namely a diamond-like pattern together with
off-centered vortex cores.
However, the weak stripe domain structure is already visible in the simulation contrary to
experiments, revealing that the computed perpendicular anisotropy is too strong compared
to the transverse term.
At still higher values, for $\Delta T =280$~K (e), a domain structure with essentially
transverse domains develops, but also with very visible weak stripes.
Thus, the observed structures are reproduced, but at the expense of large
temperature differences, resulting in a relatively too strong perpendicular
anisotropy.

\section{Conclusion and perspectives}

In this study, the magnetization distribution of NiPd nanostrips has been imaged using
complementary techniques (MFM, XMCD-PEEM and SEMPA), revealing a transverse
orientation of the magnetization and the appearance of weak stripes at low
temperature or large thickness.
The direct observation of a largely transverse magnetization differs from
the conclusions previously drawn from AMR measurements only.~\cite{Gonzalez08}
It however corresponds well with the effect of field orientation on the switching of
NiPd electrodes observed in magneto-transport measurements on carbon nanotubes,
as reported by Refs.~\onlinecite{Sahoo05a,Feuillet10}.
From these observations, it appears that in order to account for the observed textures,
non-negligible out-of-plane and transverse anisotropies have to be present.

Considering the evolution with temperature (increase of both anisotropy constants as
temperature decreases), a thermal stress mechanism has been considered as the
origin of this surprising magnetization texture, via magnetostriction.
Note that the same mechanism was invoked for explaining the spin reorientation
transition observed in Ni$_{1-x}$Pd$_x$ alloys grown on Cu$_3$Au(100).~\cite{Matthes02}
Performing elastic, magneto-elastic and micromagnetic simulations, all qualitative
features of the experiments could be reproduced, however with a disagreement
regarding the relative magnitudes of the transverse and out-of-plane anisotropies.

We conclude that in addition, another effect may be present, such
as an interfacial strain due to metal-substrate mismatch, a structural ordering of the alloy
in the growth direction (i.e., the film normal), or a plastic strain relaxation.
The latter effect may also explain the observed difference in magnetic properties
between the infinite film and the nanostructures, at the same thickness.
Indeed, weak stripes were seen to appear at lower thickness in the nanostructures,
and the value of the perpendicular anisotropy measured by ferromagnetic resonance
on infinite films was smaller than what was deduced for nanostructures of the
same thickness.
This shows also that the evaluation of strain in nanostructures is difficult, and
that magnetic patterns in nanostructures made out of magnetostrictive materials
should be imaged.

Even though only one composition has been considered in this study,
the discussion is general and should apply to other Pd
concentrations. For lower nickel concentrations, the thermal strain
is anticipated to increase, as well as the Young's modulus and
magnetostriction constant (initially at least), resulting in a
fairly constant induced anisotropy. On the other hand, the alloy
magnetization will decrease, down to zero, so that the role of the
anisotropy induced by the thermal strain will be more and more
important as the nickel content decreases. As a result, the easy
axis will switch to the direction perpendicular to the plane, at a
temperature that depends on composition. This corresponds well to
the observations at a Ni atomic concentration of
10~\%.~\cite{Kontos01}

Finally, similar phenomena should occur for nanostructures made of other materials
with a large magnetostriction and a small saturation magnetization.

\section{Acknowledgements}

We thank T. Kontos and M. Aprili for discussions, encouragements and help in the sample elaboration,
R. Weil for advice and help on sample nanofabrication, J. Ben Youssef and V. Castel
for their help in samples characterization, F. Glas for the first estimations of the
elastic deformations, and I. Vickridge at the SAFIR instrument for sample composition
measurements by RBS.
Work at LPS was partly supported by the Agence nationale de la Recherche, under
contract ANR-09-NANO-002 HYFONT.


\begin{thebibliography}{27}

\makeatletter
\providecommand \@ifxundefined [1]{%
 \@ifx{#1\undefined}
}%
\providecommand \@ifnum [1]{%
 \ifnum #1\expandafter \@firstoftwo
 \else \expandafter \@secondoftwo
 \fi
}%
\providecommand \@ifx [1]{%
 \ifx #1\expandafter \@firstoftwo
 \else \expandafter \@secondoftwo
 \fi
}%
\providecommand \natexlab [1]{#1}%
\providecommand \enquote  [1]{``#1''}%
\providecommand \bibnamefont  [1]{#1}%
\providecommand \bibfnamefont [1]{#1}%
\providecommand \citenamefont [1]{#1}%
\providecommand \href@noop [0]{\@secondoftwo}%
\providecommand \href [0]{\begingroup \@sanitize@url \@href}%
\providecommand \@href[1]{\@@startlink{#1}\@@href}%
\providecommand \@@href[1]{\endgroup#1\@@endlink}%
\providecommand \@sanitize@url [0]{\catcode `\\12\catcode `\$12\catcode
  `\&12\catcode `\#12\catcode `\^12\catcode `\_12\catcode `\%12\relax}%
\providecommand \@@startlink[1]{}%
\providecommand \@@endlink[0]{}%
\providecommand \url  [0]{\begingroup\@sanitize@url \@url }%
\providecommand \@url [1]{\endgroup\@href {#1}{\urlprefix }}%
\providecommand \urlprefix  [0]{URL }%
\providecommand \Eprint [0]{\href }%
\providecommand \doibase [0]{http://dx.doi.org/}%
\providecommand \selectlanguage [0]{\@gobble}%
\providecommand \bibinfo  [0]{\@secondoftwo}%
\providecommand \bibfield  [0]{\@secondoftwo}%
\providecommand \translation [1]{[#1]}%
\providecommand \BibitemOpen [0]{}%
\providecommand \bibitemStop [0]{}%
\providecommand \bibitemNoStop [0]{.\EOS\space}%
\providecommand \EOS [0]{\spacefactor3000\relax}%
\providecommand \BibitemShut  [1]{\csname bibitem#1\endcsname}%
\let\auto@bib@innerbib\@empty

\bibitem [{\citenamefont {Sadron}(1932)}]{Sadron32}%
  \BibitemOpen
  \bibfield  {author} {\bibinfo {author} {\bibfnamefont {C.}~\bibnamefont
  {Sadron}},\ }\href@noop {} {\bibfield  {journal} {\bibinfo  {journal} {Ann.\
  physique (Paris), Series 10,}\ }\textbf {\bibinfo {volume} {17}},\ \bibinfo
  {pages} {371} (\bibinfo {year} {1932})},\ \bibinfo {note} {{P}h. D. thesis,
  Strasbourg (1932)}\BibitemShut {NoStop}%
\bibitem [{\citenamefont {N{\'{e}}el}(1932)}]{Neel32}%
  \BibitemOpen
  \bibfield  {author} {\bibinfo {author} {\bibfnamefont {L.}~\bibnamefont
  {N{\'{e}}el}},\ }\href@noop {} {\bibfield  {journal} {\bibinfo  {journal}
  {Ann.\ physique (Paris), Series 10,}\ }\textbf {\bibinfo {volume} {17}},\
  \bibinfo {pages} {5} (\bibinfo {year} {1932})},\ \bibinfo {note} {{P}h. D.
  thesis, Strasbourg (1932)}\BibitemShut {NoStop}%
\bibitem [{\citenamefont {Crangle}\ and\ \citenamefont
  {Scott}(1965)}]{Crangle65}%
  \BibitemOpen
  \bibfield  {author} {\bibinfo {author} {\bibfnamefont {J.}~\bibnamefont
  {Crangle}}\ and\ \bibinfo {author} {\bibfnamefont {W.}~\bibnamefont
  {Scott}},\ }\href@noop {} {\bibfield  {journal} {\bibinfo  {journal} {J.\
  Appl.\ Phys.}\ }\textbf {\bibinfo {volume} {36}},\ \bibinfo {pages} {921}
  (\bibinfo {year} {1965})}\BibitemShut {NoStop}%
\bibitem [{\citenamefont {Ferrando}\ \emph {et~al.}(1972)\citenamefont
  {Ferrando}, \citenamefont {Segnan},\ and\ \citenamefont
  {Schindler}}]{Ferrando72}%
  \BibitemOpen
  \bibfield  {author} {\bibinfo {author} {\bibfnamefont {W.}~\bibnamefont
  {Ferrando}}, \bibinfo {author} {\bibfnamefont {R.}~\bibnamefont {Segnan}}, \
  and\ \bibinfo {author} {\bibfnamefont {A.}~\bibnamefont {Schindler}},\
  }\href@noop {} {\bibfield  {journal} {\bibinfo  {journal} {Phys.\ Rev.\ B}\
  }\textbf {\bibinfo {volume} {5}},\ \bibinfo {pages} {4657} (\bibinfo {year}
  {1972})}\BibitemShut {NoStop}%
\bibitem [{\citenamefont {Murani}\ \emph {et~al.}(1974)\citenamefont {Murani},
  \citenamefont {Tari},\ and\ \citenamefont {Coles}}]{Murani74}%
  \BibitemOpen
  \bibfield  {author} {\bibinfo {author} {\bibfnamefont {A.~P.}\ \bibnamefont
  {Murani}}, \bibinfo {author} {\bibfnamefont {A.}~\bibnamefont {Tari}}, \ and\
  \bibinfo {author} {\bibfnamefont {B.~R.}\ \bibnamefont {Coles}},\ }\href@noop
  {} {\bibfield  {journal} {\bibinfo  {journal} {J. Phys. F : Metal Phys.}\
  }\textbf {\bibinfo {volume} {4}},\ \bibinfo {pages} {1769} (\bibinfo {year}
  {1974})}\BibitemShut {NoStop}%
\bibitem [{\citenamefont {Kontos}\ \emph {et~al.}(2004)\citenamefont {Kontos},
  \citenamefont {Aprili}, \citenamefont {Lesueur}, \citenamefont {Grison},\
  and\ \citenamefont {Dumoulin}}]{Kontos04}%
  \BibitemOpen
  \bibfield  {author} {\bibinfo {author} {\bibfnamefont {T.}~\bibnamefont
  {Kontos}}, \bibinfo {author} {\bibfnamefont {M.}~\bibnamefont {Aprili}},
  \bibinfo {author} {\bibfnamefont {J.}~\bibnamefont {Lesueur}}, \bibinfo
  {author} {\bibfnamefont {X.}~\bibnamefont {Grison}}, \ and\ \bibinfo {author}
  {\bibfnamefont {L.}~\bibnamefont {Dumoulin}},\ }\href@noop {} {\bibfield
  {journal} {\bibinfo  {journal} {Phys.\ Rev.\ Lett.}\ }\textbf {\bibinfo
  {volume} {93}},\ \bibinfo {pages} {137001} (\bibinfo {year}
  {2004})}\BibitemShut {NoStop}%
\bibitem [{\citenamefont {Hansen}(1958)}]{Hansen58}%
  \BibitemOpen
  \bibfield  {author} {\bibinfo {author} {\bibfnamefont {M.}~\bibnamefont
  {Hansen}},\ }\href@noop {} {\emph {\bibinfo {title} {Constitution of Binary
  Alloys}}}\ (\bibinfo  {publisher} {McGraw-Hill Book Company},\ \bibinfo
  {year} {1958})\BibitemShut {NoStop}%
\bibitem [{\citenamefont {Vogel}\ \emph {et~al.}(1997)\citenamefont {Vogel},
  \citenamefont {Fontaine}, \citenamefont {Cros}, \citenamefont
  {P{\'{e}}troff}, \citenamefont {Kappler}, \citenamefont {Krill},
  \citenamefont {Rogalev},\ and\ \citenamefont {Goulon}}]{Vogel97}%
  \BibitemOpen
  \bibfield  {author} {\bibinfo {author} {\bibfnamefont {J.}~\bibnamefont
  {Vogel}}, \bibinfo {author} {\bibfnamefont {A.}~\bibnamefont {Fontaine}},
  \bibinfo {author} {\bibfnamefont {V.}~\bibnamefont {Cros}}, \bibinfo {author}
  {\bibfnamefont {F.}~\bibnamefont {P{\'{e}}troff}}, \bibinfo {author}
  {\bibfnamefont {J.-P.}\ \bibnamefont {Kappler}}, \bibinfo {author}
  {\bibfnamefont {G.}~\bibnamefont {Krill}}, \bibinfo {author} {\bibfnamefont
  {A.}~\bibnamefont {Rogalev}}, \ and\ \bibinfo {author} {\bibfnamefont
  {J.}~\bibnamefont {Goulon}},\ }\href@noop {} {\bibfield  {journal} {\bibinfo
  {journal} {Phys.\ Rev.\ B}\ }\textbf {\bibinfo {volume} {55}},\ \bibinfo
  {pages} {3663} (\bibinfo {year} {1997})}\BibitemShut {NoStop}%
\bibitem [{\citenamefont {Kontos}\ \emph {et~al.}(2001)\citenamefont {Kontos},
  \citenamefont {Aprili}, \citenamefont {Lesueur},\ and\ \citenamefont
  {Grison}}]{Kontos01}%
  \BibitemOpen
  \bibfield  {author} {\bibinfo {author} {\bibfnamefont {T.}~\bibnamefont
  {Kontos}}, \bibinfo {author} {\bibfnamefont {M.}~\bibnamefont {Aprili}},
  \bibinfo {author} {\bibfnamefont {J.}~\bibnamefont {Lesueur}}, \ and\
  \bibinfo {author} {\bibfnamefont {X.}~\bibnamefont {Grison}},\ }\href@noop {}
  {\bibfield  {journal} {\bibinfo  {journal} {Phys.\ Rev.\ Lett.}\ }\textbf
  {\bibinfo {volume} {86}},\ \bibinfo {pages} {304} (\bibinfo {year}
  {2001})}\BibitemShut {NoStop}%
\bibitem [{\citenamefont {Javey}\ \emph {et~al.}(2003)\citenamefont {Javey},
  \citenamefont {Guo}, \citenamefont {Wang}, \citenamefont {Lundstrom},\ and\
  \citenamefont {Dai}}]{Javey03}%
  \BibitemOpen
  \bibfield  {author} {\bibinfo {author} {\bibfnamefont {A.}~\bibnamefont
  {Javey}}, \bibinfo {author} {\bibfnamefont {J.}~\bibnamefont {Guo}}, \bibinfo
  {author} {\bibfnamefont {Q.}~\bibnamefont {Wang}}, \bibinfo {author}
  {\bibfnamefont {M.}~\bibnamefont {Lundstrom}}, \ and\ \bibinfo {author}
  {\bibfnamefont {H.}~\bibnamefont {Dai}},\ }\href@noop {} {\bibfield
  {journal} {\bibinfo  {journal} {Nature}\ }\textbf {\bibinfo {volume} {424}},\
  \bibinfo {pages} {654} (\bibinfo {year} {2003})}\BibitemShut {NoStop}%
\bibitem [{\citenamefont {Sahoo}\ \emph
  {et~al.}(2005{\natexlab{a}})\citenamefont {Sahoo}, \citenamefont {Kontos},
  \citenamefont {Sch{\"o}nenberger},\ and\ \citenamefont
  {S{\"u}rgers}}]{Sahoo05a}%
  \BibitemOpen
  \bibfield  {author} {\bibinfo !a!L {author} {\bibfnamefont {S.}~\bibnamefont
  {Sahoo}}, \bibinfo {author} {\bibfnamefont {T.}~\bibnamefont {Kontos}},
  \bibinfo {author} {\bibfnamefont {C.}~\bibnamefont {Sch{\"o}nenberger}}, \
  and\ \bibinfo {author} {\bibfnamefont {C.}~\bibnamefont {S{\"u}rgers}},\
  }\href@noop {} {\bibfield  {journal} {\bibinfo  {journal} {Appl.\ Phys.\
  Lett.}\ }\textbf {\bibinfo {volume} {86}},\ \bibinfo {pages} {112109}
  (\bibinfo {year} {2005}{\natexlab{a}})}\BibitemShut {NoStop}%
\bibitem [{\citenamefont {Sahoo}\ \emph
  {et~al.}(2005{\natexlab{b}})\citenamefont {Sahoo}, \citenamefont {Kontos},
  \citenamefont {Furer}, \citenamefont {Hoffmann}, \citenamefont {Gr{\"a}ber},
  \citenamefont {Cottet},\ and\ \citenamefont {Sch{\"o}nenberger}}]{Sahoo05b}%
  \BibitemOpen
  \bibfield  {author} {\bibinfo {author} {\bibfnamefont {S.}~\bibnamefont
  {Sahoo}}, \bibinfo {author} {\bibfnamefont {T.}~\bibnamefont {Kontos}},
  \bibinfo {author} {\bibfnamefont {J.}~\bibnamefont {Furer}}, \bibinfo
  {author} {\bibfnamefont {C.}~\bibnamefont {Hoffmann}}, \bibinfo {author}
  {\bibfnamefont {M.}~\bibnamefont {Gr{\"a}ber}}, \bibinfo {author}
  {\bibfnamefont {A.}~\bibnamefont {Cottet}}, \ and\ \bibinfo {author}
  {\bibfnamefont {C.}~\bibnamefont {Sch{\"o}nenberger}},\ }\href@noop {}
  {\bibfield  {journal} {\bibinfo  {journal} {Nature Phys.}\ }\textbf {\bibinfo
  {volume} {1}},\ \bibinfo {pages} {99} (\bibinfo {year}
  {2005}{\natexlab{b}})}\BibitemShut {NoStop}%
\bibitem [{\citenamefont {Man}\ \emph {et~al.}(2006)\citenamefont {Man},
  \citenamefont {Wever},\ and\ \citenamefont {Morpurgo}}]{Man06}%
  \BibitemOpen
  \bibfield  {author} {\bibinfo {author} {\bibfnamefont {H.~T.}\ \bibnamefont
  {Man}}, \bibinfo {author} {\bibfnamefont {I.~J.~W.}\ \bibnamefont {Wever}}, \
  and\ \bibinfo {author} {\bibfnamefont {A.~F.}\ \bibnamefont {Morpurgo}},\
  }\href@noop {} {\bibfield  {journal} {\bibinfo  {journal} {Phys.\ Rev.\ B}\
  }\textbf {\bibinfo {volume} {73}},\ \bibinfo {pages} {241401(R)} (\bibinfo
  {year} {2006})}\BibitemShut {NoStop}%
\bibitem [{\citenamefont {Feuillet-Palma}\ \emph {et~al.}(2010)\citenamefont
  {Feuillet-Palma}, \citenamefont {Delattre}, \citenamefont {Morfin},
  \citenamefont {Berroir}, \citenamefont {Feve}, \citenamefont {Glattli},
  \citenamefont {Pla{\c{c}}ais}, \citenamefont {Cottet},\ and\ \citenamefont
  {Kontos}}]{Feuillet10}%
  \BibitemOpen
  \bibfield  {author} {\bibinfo {author} {\bibfnamefont {C.}~\bibnamefont
  {Feuillet-Palma}}, \bibinfo {author} {\bibfnamefont {T.}~\bibnamefont
  {Delattre}}, \bibinfo {author} {\bibfnamefont {P.}~\bibnamefont {Morfin}},
  \bibinfo {author} {\bibfnamefont {J.~M.}\ \bibnamefont {Berroir}}, \bibinfo
  {author} {\bibfnamefont {G.}~\bibnamefont {Feve}}, \bibinfo {author}
  {\bibfnamefont {D.~C.}\ \bibnamefont {Glattli}}, \bibinfo {author}
  {\bibfnamefont {B.}~\bibnamefont {Pla{\c{c}}ais}}, \bibinfo {author}
  {\bibfnamefont {A.}~\bibnamefont {Cottet}}, \ and\ \bibinfo {author}
  {\bibfnamefont {T.}~\bibnamefont {Kontos}},\ }\href@noop {} {\bibfield
  {journal} {\bibinfo  {journal} {Phys.\ Rev.\ B}\ }\textbf {\bibinfo {volume}
  {81}},\ \bibinfo {pages} {115414} (\bibinfo {year} {2010})}\BibitemShut
  {NoStop}%
\bibitem [{\citenamefont {Gonzalez-Pons}\ \emph {et~al.}(2008)\citenamefont
  {Gonzalez-Pons}, \citenamefont {Henderson}, \citenamefont {del Barco},\ and\
  \citenamefont {Ozyilmaz}}]{Gonzalez08}%
  \BibitemOpen
  \bibfield  {author} {\bibinfo {author} {\bibfnamefont {J.~C.}\ \bibnamefont
  {Gonzalez-Pons}}, \bibinfo {author} {\bibfnamefont {J.~J.}\ \bibnamefont
  {Henderson}}, \bibinfo {author} {\bibfnamefont {E.}~\bibnamefont {del
  Barco}}, \ and\ \bibinfo {author} {\bibfnamefont {B.}~\bibnamefont
  {Ozyilmaz}},\ }\href@noop {} {\bibfield  {journal} {\bibinfo  {journal}
  {Phys.\ Rev.\ B}\ }\textbf {\bibinfo {volume} {78}},\ \bibinfo {pages}
  {012408} (\bibinfo {year} {2008})}\BibitemShut {NoStop}%
\bibitem [{\citenamefont {Hopster}\ and\ \citenamefont
  {Oepen}(2005)}]{Hopster05}%
  \BibitemOpen
  \bibfield  {author} {\bibinfo {author} {\bibfnamefont {H.}~\bibnamefont
  {Hopster}}\ and\ \bibinfo {author} {\bibfnamefont {H.~P.}\ \bibnamefont
  {Oepen}},\ }\href@noop {} {\emph {\bibinfo {title} {Magnetic microscopy of
  nanostructures}}}\ (\bibinfo  {publisher} {Springer Verlag},\ \bibinfo
  {address} {Berlin},\ \bibinfo {year} {2005})\BibitemShut {NoStop}%
\bibitem [{\citenamefont {Locatelli}\ \emph {et~al.}(2006)\citenamefont
  {Locatelli}, \citenamefont {Aballe}, \citenamefont {Mentes}, \citenamefont
  {Kiskinova},\ and\ \citenamefont {Bauer}}]{Locatelli06}%
  \BibitemOpen
  \bibfield  {author} {\bibinfo {author} {\bibfnamefont {A.}~\bibnamefont
  {Locatelli}}, \bibinfo {author} {\bibfnamefont {L.}~\bibnamefont {Aballe}},
  \bibinfo {author} {\bibfnamefont {T.}~\bibnamefont {Mentes}}, \bibinfo
  {author} {\bibfnamefont {M.}~\bibnamefont {Kiskinova}}, \ and\ \bibinfo
  {author} {\bibfnamefont {E.}~\bibnamefont {Bauer}},\ }\href@noop {}
  {\bibfield  {journal} {\bibinfo  {journal} {Surf.\ Interface\ Anal.}\
  }\textbf {\bibinfo {volume} {38}},\ \bibinfo {pages} {1554} (\bibinfo {year}
  {2006})}\BibitemShut {NoStop}%
\bibitem [{\citenamefont {Hubert}\ and\ \citenamefont
  {Sch{{\"a}}fer}(1998)}]{Hubert98-WS}%
  \BibitemOpen
  \bibfield  {author} {\bibinfo {author} {\bibfnamefont {A.}~\bibnamefont
  {Hubert}}\ and\ \bibinfo {author} {\bibfnamefont {R.}~\bibnamefont
  {Sch{{\"a}}fer}},\ }\href@noop {} {\emph {\bibinfo {title} {Magnetic
  Domains}}}\ (\bibinfo  {publisher} {Springer Verlag},\ \bibinfo {address}
  {Berlin},\ \bibinfo {year} {1998})\ pp.\ \bibinfo {pages}
  {298--303}\BibitemShut {NoStop}%
\bibitem [{\citenamefont {Yamamoto}(1951)}]{Yamamoto51}%
  \BibitemOpen
  \bibfield  {author} {\bibinfo {author} {\bibfnamefont {M.}~\bibnamefont
  {Yamamoto}},\ }\href@noop {} {\bibfield  {journal} {\bibinfo  {journal}
  {Science reports of the research institutes, Tohoku Univ., Ser. A}\ }\textbf
  {\bibinfo {volume} {3}},\ \bibinfo {pages} {308} (\bibinfo {year} {1951})},\
  \bibinfo {note} {available at http://hdl.handle.net/10097/26440}\BibitemShut
  {NoStop}%
\bibitem [{\citenamefont {Rayne}(1960)}]{Rayne60}%
  \BibitemOpen
  \bibfield  {author} {\bibinfo {author} {\bibfnamefont {J.}~\bibnamefont
  {Rayne}},\ }\href@noop {} {\bibfield  {journal} {\bibinfo  {journal} {Phys.\
  Rev.}\ }\textbf {\bibinfo {volume} {118}},\ \bibinfo {pages} {1545} (\bibinfo
  {year} {1960})}\BibitemShut {NoStop}%
\bibitem [{\citenamefont {{Ben Youssef}}\ \emph {et~al.}(2004)\citenamefont
  {{Ben Youssef}}, \citenamefont {Vukadinovic}, \citenamefont {Billet},\ and\
  \citenamefont {Labrune}}]{BenYoussef04}%
  \BibitemOpen
  \bibfield  {author} {\bibinfo {author} {\bibfnamefont {J.}~\bibnamefont {{Ben
  Youssef}}}, \bibinfo {author} {\bibfnamefont {N.}~\bibnamefont
  {Vukadinovic}}, \bibinfo {author} {\bibfnamefont {D.}~\bibnamefont {Billet}},
  \ and\ \bibinfo {author} {\bibfnamefont {M.}~\bibnamefont {Labrune}},\
  }\href@noop {} {\bibfield  {journal} {\bibinfo  {journal} {Phys.\ Rev.\ B}\
  }\textbf {\bibinfo {volume} {69}},\ \bibinfo {pages} {174402} (\bibinfo
  {year} {2004})}\BibitemShut {NoStop}%
\bibitem [{\citenamefont {Wortman}\ and\ \citenamefont
  {Evans}(1965)}]{Wortman65}%
  \BibitemOpen
  \bibfield  {author} {\bibinfo {author} {\bibfnamefont {J.}~\bibnamefont
  {Wortman}}\ and\ \bibinfo {author} {\bibfnamefont {R.}~\bibnamefont
  {Evans}},\ }\href@noop {} {\bibfield  {journal} {\bibinfo  {journal} {J.\
  Appl.\ Phys.}\ }\textbf {\bibinfo {volume} {36}},\ \bibinfo {pages} {153}
  (\bibinfo {year} {1965})}\BibitemShut {NoStop}%
\bibitem [{\citenamefont {Vukadinovic}\ \emph {et~al.}(2001)\citenamefont
  {Vukadinovic}, \citenamefont {Labrune}, \citenamefont {{Ben Youssef}},
  \citenamefont {Marty}, \citenamefont {Toussaint},\ and\ \citenamefont {{Le
  Gall}}}]{Vukadinovic01}%
  \BibitemOpen
  \bibfield  {author} {\bibinfo {author} {\bibfnamefont {N.}~\bibnamefont
  {Vukadinovic}}, \bibinfo {author} {\bibfnamefont {M.}~\bibnamefont
  {Labrune}}, \bibinfo {author} {\bibfnamefont {J.}~\bibnamefont {{Ben
  Youssef}}}, \bibinfo {author} {\bibfnamefont {A.}~\bibnamefont {Marty}},
  \bibinfo {author} {\bibfnamefont {J.~C.}\ \bibnamefont {Toussaint}}, \ and\
  \bibinfo {author} {\bibfnamefont {H.}~\bibnamefont {{Le Gall}}},\ }\href@noop
  {} {\bibfield  {journal} {\bibinfo  {journal} {Phys.\ Rev.\ B}\ }\textbf
  {\bibinfo {volume} {65}},\ \bibinfo {pages} {054403} (\bibinfo {year}
  {2001})}\BibitemShut {NoStop}%
\bibitem [{\citenamefont {Masumoto}\ and\ \citenamefont
  {Sawaya}(1970)}]{Masumoto70}%
  \BibitemOpen
  \bibfield  {author} {\bibinfo {author} {\bibfnamefont {H.}~\bibnamefont
  {Masumoto}}\ and\ \bibinfo {author} {\bibfnamefont {S.}~\bibnamefont
  {Sawaya}},\ }\href@noop {} {\bibfield  {journal} {\bibinfo  {journal}
  {Trans.\ J.\ I.\ M.}\ }\textbf {\bibinfo {volume} {11}},\ \bibinfo {pages}
  {391} (\bibinfo {year} {1970})}\BibitemShut {NoStop}%
\bibitem [{\citenamefont {Tokunaga}\ and\ \citenamefont
  {Fujiwara}(1978)}]{Tokunaga78}%
  \BibitemOpen
  \bibfield  {author} {\bibinfo {author} {\bibfnamefont {T.}~\bibnamefont
  {Tokunaga}}\ and\ \bibinfo {author} {\bibfnamefont {H.}~\bibnamefont
  {Fujiwara}},\ }\href@noop {} {\bibfield  {journal} {\bibinfo  {journal} {J.\
  Phys.\ Soc.\ Japan}\ }\textbf {\bibinfo {volume} {45}},\ \bibinfo {pages}
  {1232} (\bibinfo {year} {1978})}\BibitemShut {NoStop}%
\bibitem [{OOM()}]{OOMMF}%
  \BibitemOpen
  \href@noop {} {}\bibinfo {note} {OOMMF is a free software (in fact, an open
  framework for micromagnetics routines) developped by M.J. Donahue and D.
  Porter mainly, from NIST. It is available at
  http://math.nist.gov/oommf.}\BibitemShut {Stop}%
\bibitem [{\citenamefont {Matthes}\ \emph {et~al.}(2002)\citenamefont
  {Matthes}, \citenamefont {Seider},\ and\ \citenamefont
  {Schneider}}]{Matthes02}%
  \BibitemOpen
  \bibfield  {author} {\bibinfo {author} {\bibfnamefont {F.}~\bibnamefont
  {Matthes}}, \bibinfo {author} {\bibfnamefont {M.}~\bibnamefont {Seider}}, \
  and\ \bibinfo {author} {\bibfnamefont {C.}~\bibnamefont {Schneider}},\
  }\href@noop {} {\bibfield  {journal} {\bibinfo  {journal} {J.\ Appl.\ Phys.}\
  }\textbf {\bibinfo {volume} {91}},\ \bibinfo {pages} {8144} (\bibinfo {year}
  {2002})}\BibitemShut {NoStop}%
\end{thebibliography}
\end{document}